\documentclass{article}
\usepackage{spconf,amsmath,graphicx,booktabs}
\usepackage{multirow}
\usepackage{makecell}
\usepackage{amsfonts}
\usepackage{verbatim}
\usepackage{threeparttable}
\usepackage{booktabs}
\usepackage{color}
\usepackage{subfigure}

\title{A comparison of handcrafted, parameterized, and learnable features for speech separation}
\name{Wenbo Zhu, Mou Wang, Xiao-Lei Zhang, Susanto Rahardja}

\address{CIAIC, School of Marine Science and Technology, Northwestern Polytechnical University, China}

\email{\{wbzhu, wangmou21\}@mail.nwpu.edu.cn,
\{xiaolei.zhang, susanto\}@nwpu.edu.cn}
\begin{document}
%
\maketitle
\begin{abstract}
The design of acoustic features is important for speech separation. It can be roughly categorized into three classes: handcrafted, parameterized, and learnable features. Among them, learnable features, which are trained with separation networks jointly in an end-to-end fashion, become a new trend of modern speech separation research, e.g. convolutional time domain audio separation network (Conv-Tasnet), while handcrafted and parameterized features are also shown competitive in very recent studies. However, a systematic comparison across the three kinds of acoustic features has not been conducted yet. In this paper, we compare them in the framework of Conv-Tasnet by setting its encoder and decoder with different acoustic features. We also generalize the handcrafted multi-phase gammatone filterbank (MPGTF) to a new parameterized multi-phase gammatone filterbank (ParaMPGTF).  Experimental results on the WSJ0-2mix corpus show that (i) if the decoder is learnable, then setting the encoder to STFT, MPGTF, ParaMPGTF, and learnable features lead to similar performance; and (ii) when the pseudo-inverse transforms of STFT, MPGTF, and ParaMPGTF are used as the decoders, the proposed ParaMPGTF performs better than the other two handcrafted features.
\end{abstract}
\section{Introduction}

Speech separation aims to separate a mixture of multiple speech sources into its components.
In this paper, we study deep learning based speaker-independent speech separation, which does not require training and test speakers to be the same \cite{8369155}. Hershey \textit{et al.} first addressed the problem by deep clustering \cite{7471631}. Since then, several methods have been proposed, such as permutation invariant training \cite{7952154,7979557} and deep attractor networks \cite{7952155} which aim to estimate a time-frequency mask for each speaker. Among the methods, the magnitude spectrogram of short-time Fourier transform (STFT) is the most widely used acoustic feature. However, when recovering the time-domain speech from the separated magnitude spectrograms, the noisy phase has to be used, which results in suboptimal performance.

To remedy this weakness, learnable features, which learn a network for the transforms between the time-domain signal and its time-frequency spectrogram are becoming a new trend. Representative ones include one-dimensional convolution (1D-conv) filters \cite{1711.00541,1809.07454}\cite{8683634,1902.04891}. Because the transforms are jointly trained with the separation network, and also because they do not need additional handcrafted operations, they lead to improved performance over STFT. Among the time-domain speech separation methods, convolutional time domain audio separation network (Conv-Tasnet), which reaches outstanding separation performance with a frame length of only 2 ms, received much attention.

Several recent work studied acoustic features with Conv-Tasnet. For example, Ditter and Gerkmann \cite{9053602} used a handcrafted feature, named
 multi-phase gammatone filterbank (MPGTF), to replace the 1D-conv learnable feature of the encoder, which leads to improvement over the original Conv-Tasnet in terms of the scale-invariant source-to-noise ratio (SI-SNR). Pariente \textit{et al.} \cite{1910.10400} extended the parameterized filters introduced in \cite{1808.00158} to complex-valued analytic filters, and then proposed a similar analytic extension for the 1D-conv filter as well. The analytic 1D-conv filter improves the performance as well. The aforementioned positive results demonstrate that handcrafted and parameterized features are also competitive to the state-of-the-art learnable features.

 However, there lacks a comparison between the handcrafted, parameterized, and learnable features. Motivated by replacing the encoder or decoder by handcrafted features, in this paper, we compared the three kinds of features in the framework of Conv-Tasnet. To understand the connection between the three kinds of features, we proposed a parameterized extension of MPGTF, named parameterized MPGTF (ParaMPGTF). The center frequencies and bandwidths of ParaMPGTF are jointly trained with the separation network.
 We conducted an experimental comparison between STFT, MPGTF, ParaMPGTF, and learnable features on WSJ0-2mix \cite{7471631}. Experimental results show that, if the decoder is learnable, then setting the encoder to any of the comparison features leads to similar performance. We have also compared STFT, MPGTF, and ParaMPGTF when their (pseudo) inverse transforms are used as the decoders. Results show that the proposed ParaMPGTF performs better than the other two handcrafted features.

This paper is organized as follows. Section \ref{sec:model} presents the comparison framework and the proposed ParaMPGTF. Section \ref{sec:experiment} presents the experimental results. Finally, we conclude our findings in Section \ref{sec:conclusion}.

\section{Methods}
\label{sec:model}

\begin{figure}
    \centering
    \includegraphics[width=0.45\textwidth]{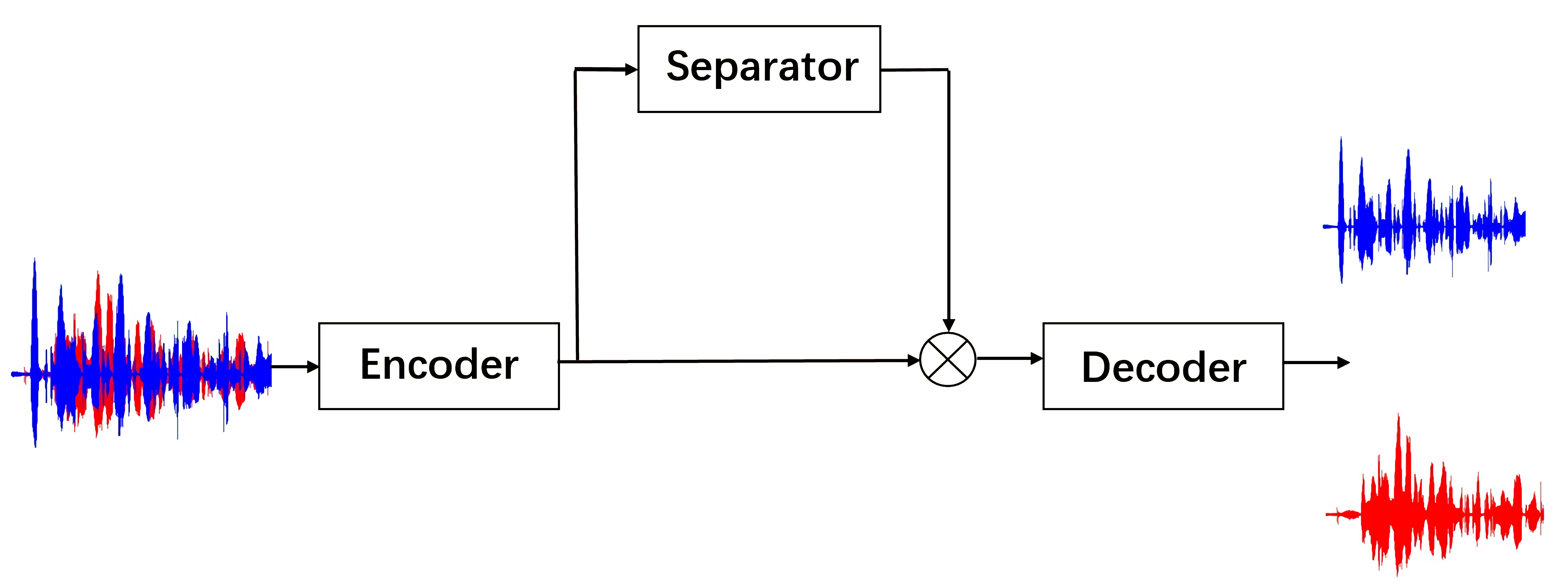}
    \caption{The building blocks of Conv-Tasnet}
    \label{Conv-Tasnet_jpg}
\end{figure}

\subsection{Preliminary}

Given $C$ speech sources $\left\{\mathbf{s}_{c}(t)\right\}_{c=1}^C$ with $t$ as the index of time samples, their mixed signal is
\begin{equation}
\mathbf{x}(t)=\sum_{c=1}^{C} \mathbf{s}_{c}(t)
\end{equation}
The problem of speech separation can be described as producing an accurate estimate $\hat{\mathbf{s}}_{c}(t)$ for $\mathbf{s}_{c}(t)$ from $\mathbf{x}(t)$.

The framework in this study is Conv-Tasnet \cite{1809.07454}. As shown in Fig. \ref{Conv-Tasnet_jpg}, it consists of three main parts---an encoder, a separation network, and a decoder. It uses a small frame size in the encoder and decoder to reduce the time delay significantly. The encoder and decoder are learnable 1D-conv filters, which perform like transforms between the time-domain signal and time-frequency features.
The separation network is a fully-convolutional separation module that consists of stacked one-dimensional dilated convolutional blocks \cite{10.1007/978-3-319-49409-8_7,8099596}. It is optimized with the scale-invariant soure-to-noise ratio (SI-SNR) loss \cite{7952155}. It produces a mask for each speech source.

\subsection{Comparison framework}

The comparison uses handcrafted transforms, parameterized transforms, and learnable filters as the encoder and decoder. The encoder can be thought of as s set of $N$ filters of length $L$. The output of the encoder is a time-frequency representation produced from the convolution of the mixed speech input signal with the filter:
\begin{equation}
\mathbf{X}(n, i)=\mathcal{H}(\sum_{l=0}^{L-1} \mathbf{x}(i D+l) \mathbf{h}_{n}^{\mathrm{Enc}}(L-l))
\end{equation}
where $n$ is the filter index, $i$ is the frame index, $D$ is the frame shift, $\mathbf{h}_{n}^{Enc}(\cdot)$ is the $n$-th filter of the filterbanks, $l$ denotes the sample index in a frame, and $\mathcal{H}(\cdot)$ is the rectified linear unit (ReLU) to ensure that the representation is non-negative. In the comparison, $\mathbf{h}_{n}^{\mathrm{Enc}}(\cdot)$ can be any of the three kinds of feature transforms.

 The decoder reconstructs the time-domain signal of the $c$-th speaker $\hat{\mathbf{s}}_{c} \in \mathbb{R}^{T}$. The output of the decoder is:
\begin{equation}
\hat{\mathbf{s}}_{c}(k, i)=\sum_{n=0}^{N-1} \hat{\mathbf{S}}_{c}(n, i) \mathbf{h}_{N-n}^{\mathrm{Dec}}(k)
\end{equation}
where $ \hat{\mathbf{S}}_{c}(n, i)$ is the output of the separation network for the $c$-th speaker, $k$ is the index of the filter weight, $\mathbf{h}_{n}^{\rm Dec}(\cdot) $ is the $n$-th filter of the decoder, and $\hat{\mathbf{s}}_{c}(k, i)$ is the estimate of the $c$-th speech source at the $i$-th frame. To decode the frame-shift operation between speech frames, the decoder further calculates $\hat{\mathbf{s}}_{c}(t)=\sum_{i=-\infty}^{\infty} \hat{\mathbf{s}}_{c}(t-i D, i)$.

The comparison uses STFT, MPGTF, ParaMPGTF, and learnable filters as $\mathbf{h}_{n}^{\mathrm{Enc}}(\cdot)$ with their inverse transforms as $\mathbf{h}_{N-n}^{\mathrm{Dec}}(\cdot)$, where the proposed ParaMPGTF is presented in the next subsection.

\subsection{Parameterized multi-phase gammatone filterbank}

Gammatone filterbank, which mimics the masking effect of the human auditory system, are good features for speech separation \cite{1992Complex}. The impulse response function $\gamma(t)$ of a gammatone filter is
\begin{equation}\label{eq:xxx}
\gamma(t)=\alpha t^{n-1}\exp(-2\pi bt)\cos(2\pi f_{c}t+\phi)
\end{equation}
where $n$ is the order, $b$ is a bandwidth parameter, $f_{c}$ is the centre frequency of the filter, $t \textgreater 0$ is the time in seconds, $\alpha$ is the amplitude, and $\phi$ is the phase shift.
Ditter and Gerkmann \cite{9053602} extended the classical gammatone filterbank to MPGTF. The extension has the following three three aspects: First, the length of the filters is set to $2$ms, which keeps the system a low latency. Second, for each filter $h_n^{\rm Enc}(\cdot)$, MPGTF introduces $-h_{n}^{\rm Enc}(\cdot)$ to ensure that, for each centre frequency, at least one filter contains energy. Third, the phase shift $\phi$ varies at the same centre frequency. The details of MPGTF can be found in \cite{9053602}.

From \eqref{eq:xxx}, we observe that the bandwidth parameter $b$ and filter centre frequency $f_{c}$ are two important parameters.
They are determined by the equivalent rectangular bandwidth (ERB) \cite{2002Frequency} using a rectangular band-pass filter:
  \begin{eqnarray}
  	\label{para_ERB}
  	&&\mathrm{ERB}(f_{c},c_{1},c_{2})=c_{1}+\frac{f_{c}}{c_2}\label{eq:1}\\
 &&\textcolor[rgb]{0.00,0.07,1.00}{ f_{c}= c_2 (\mathrm{ERB}-c_1)\label{eq:2}}\\
 && b=\frac{\mathrm{ERB}\sqrt{(n-1)!}}{\pi \left((2n-2)!\right)2^{2-2n}}\label{eq:3}
  \end{eqnarray}
 where $c_{1}$ and $c_{2}$ are two parameters.
 Traditionally, the parameters $c_{1}$ and $c_{2}$ are set to $24.7$ and $9.265$ respectively in experience \cite{2002Frequency}. This empirical setting may not be accurate enough, which may lead to suboptimal performance.

To overcome this issue, we propose ParaMPGTF which trains the filterbank parameters $c_{1}$ and $c_{2}$ in MPGTF jointly with the network. For each iteration, we update the parameter $b$ by \eqref{eq:3} and the centre frequencies $f_{c_1},f_{c_2},\dots, f_{c_M}$ by:
\begin{equation}
\label{fc_iter}
f_{c_j}=\mathrm{ERB}_{\rm scale}^{-1}(\mathrm{ERB}_{\rm scale}(f_{c_{j-1}})+1)
\end{equation}
according to the updated $c_{1}$ and $c_{2}$,
where $f_{c_j}$ denotes the centre frequency of the $j$-th filter, $M$ is the number of filters in the  filterbank, $\mathrm{ERB}_{\rm scale}$ denotes the ERB scale calculated by integrating $1/\mathrm{ERB}(f_{c})$ across frequency, and $\mathrm{ERB}^{-1}_{\rm scale}$ is the inverse of $\mathrm{ERB}_{\rm scale}$. In practice, $\mathrm{ERB}_{\rm scale}$ and $\mathrm{ERB}_{\rm scale}^{-1}$ are calculated by:
\begin{eqnarray}
	&&\mathrm{ERB}_{\rm scale}(f_{\mathrm{Hz}})=c_{2} \log(1+\frac{f_{\mathrm{Hz}}}{c_{1} c_{2}})\\
	&&\mathrm{ERB}^{-1}_{\rm scale}(\mathrm{ERB}_{\rm scale})=c_{1}c_{2}(\mathrm{e^{{\frac{\mathrm{ERB}_{\rm scale}}{c_{2}}}}}-1)
\end{eqnarray}
where $f_{\mathrm{Hz}}$ denotes a frequency variable. After obtaining $f_{c_1},\dots, f_{c_M}$ and $b$, we obtain the updated filterbank according to \eqref{eq:xxx}. To make ParaMPGTF a meaningful filterbank, $f_{c_1},f_{c_2},\dots, f_{c_M}$ should be constrained between $100$ Hz to $4000$ Hz. To satisfy this constraint, we fix $f_{c_1}$ to $100$ Hz in the entire training process.

To summarize, ParaMPGTF combines the data-driven scheme with MPGTF \cite{9053602}. It inherits the changes of MPGTF.

\section{EXPERIMENTALS AND RESULTS}
\label{sec:experiment}
\subsection{Dataset}

We conducted the comparison on two-speaker speech separation using the WSJ0-2mix dataset \cite{7471631}. It contains $30$ hours training data, $10$ hours development data, and $5$ hours test data. The mixtures in WSJ0-2mix were generated by first randomly selecting different speakers and utterances in the Wall Street Journal (WSJ0) training set si\_tr\_s, and then mixing them at a random signal-to-noise ratio (SNR) level between -$5$ dB and $5$ dB \cite{1809.07454}. The utterances in the test set were from $16$ unseen speakers in the si\_dt\_05 and si\_et\_05 directories of the WSJ0 dataset. All waveforms were resampled to $8$ kHz.

\subsection{Experimental setup}
The network was trained for $200$ epochs on $4$-second long segments. Adam was used as the optimizer with an initial learning rate of $0.001$. The learning rate was halved if the performance of development set was not improved in $5$ consecutive epochs. The network training procedure was early stopped when the performance on the development set has not been improved within the last $10$ epochs. The hyperparameters of the network followed the setting in \cite{9053602}, where the number of filters $N$ is 512. The mask activation function of TCN was set to sigmoid function and rectified linear unit (ReLU) respectively.

For ParaMPGTF, we set the order $n$ and amplitude $\alpha$  to 2 and 1 respectively. We initialize $c_{1}$ and $c_{2}$ to their empirical values, i.e. $c_{1} = 24.7$ and $c_{2} =9.265$.

We used SI-SNR as the evaluation metric \cite{7952155}. We reported the average results over all $3000$ test mixtures.



\begin{figure*}[ht]
\centering
\subfigure[MPGTF-Learned]{
\begin{minipage}{5.5cm}
\centering
\label{MPGTF_learned_subfig}
    \includegraphics[width=5.2cm,height=5.9cm]{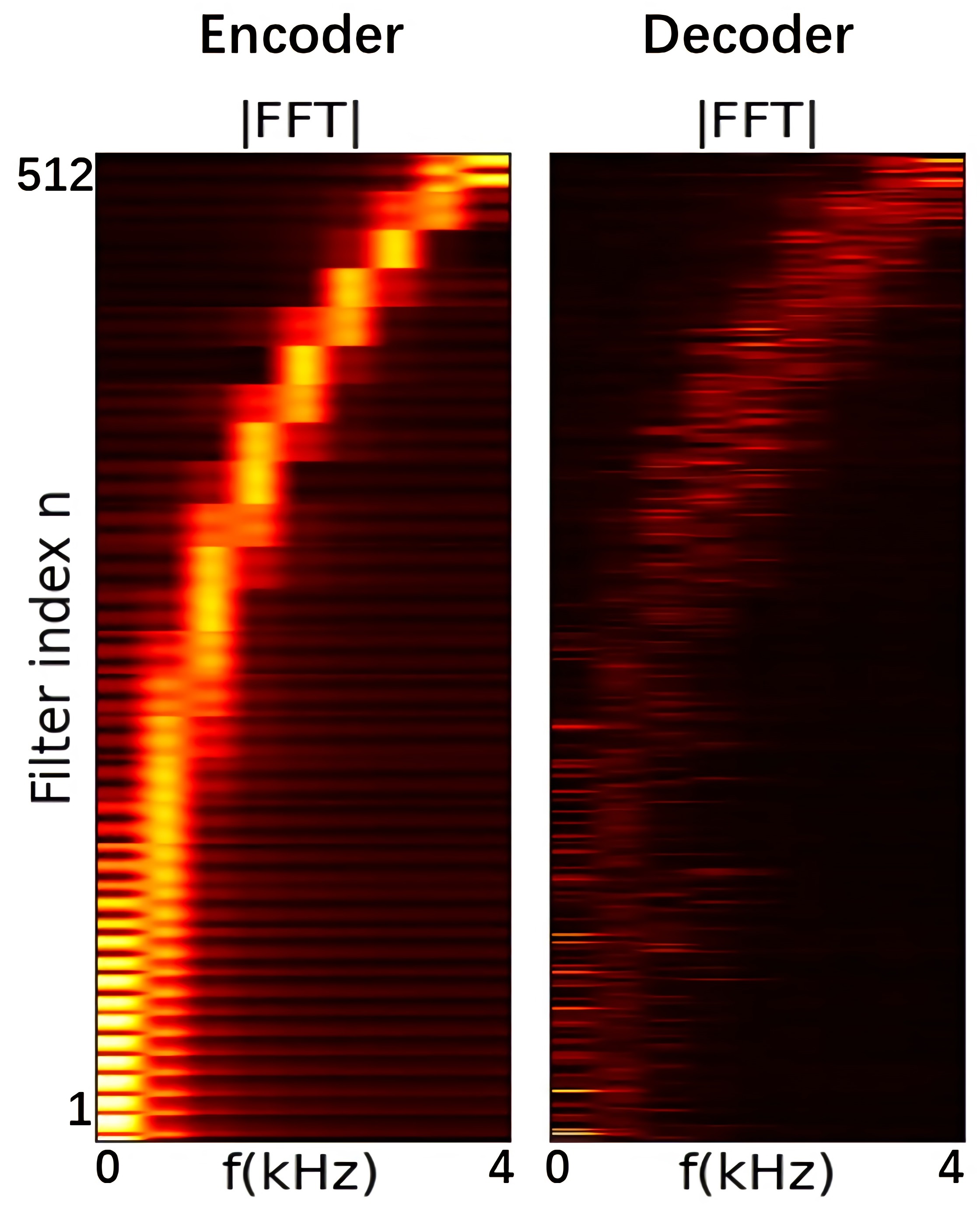}
\end{minipage}
}
\subfigure[ParaMPGTF-Learned]{
\begin{minipage}{5.5cm}
\centering
\label{ParaMPGTF_learned_subfig}
    \includegraphics[width=5.2cm,height=5.9cm]{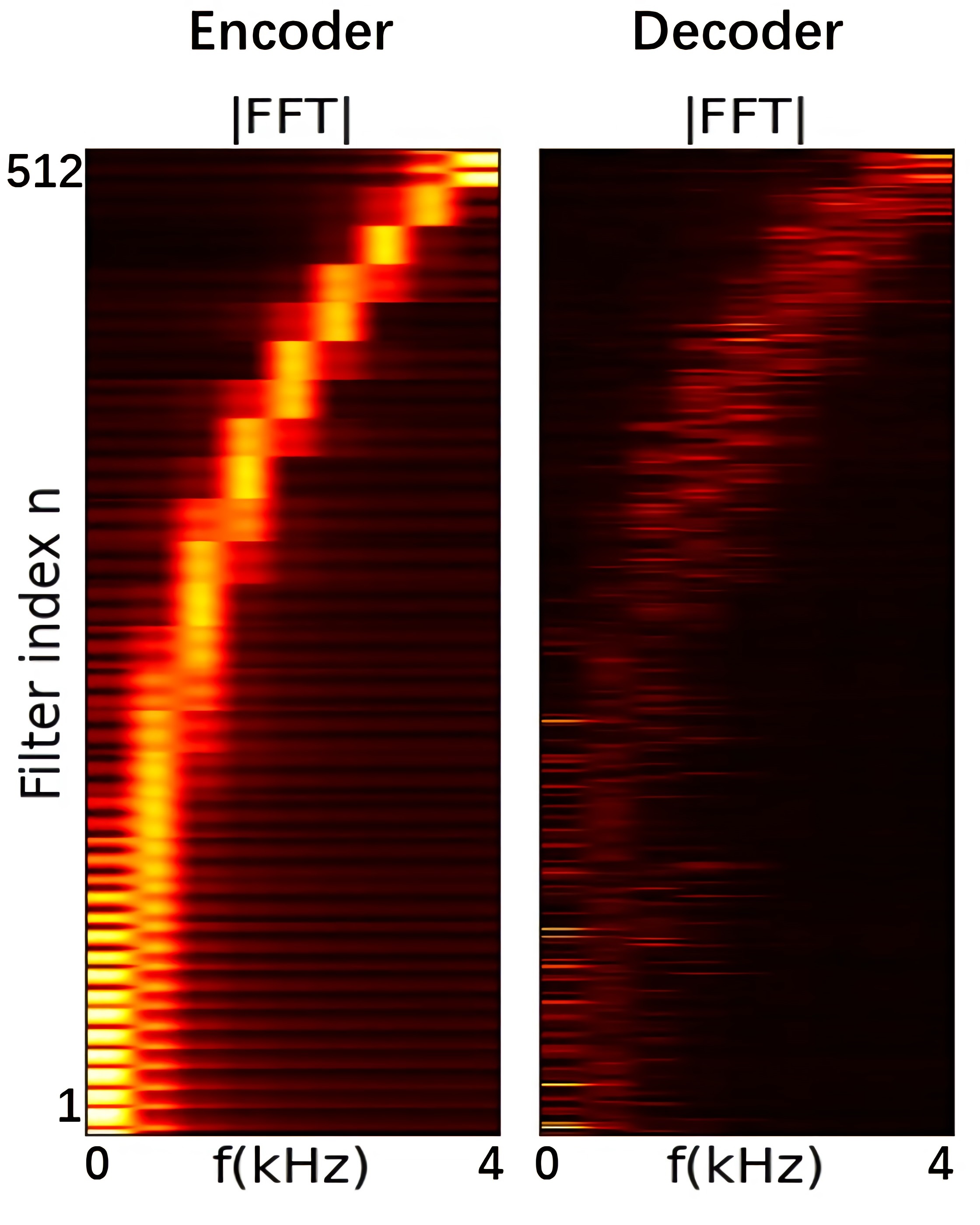}
\end{minipage}
}
\subfigure[STFT-Learned]{
\begin{minipage}{5.5cm}
\centering
\label{STFT_learned_subfig}
    \includegraphics[width=5.2cm,height=5.9cm]{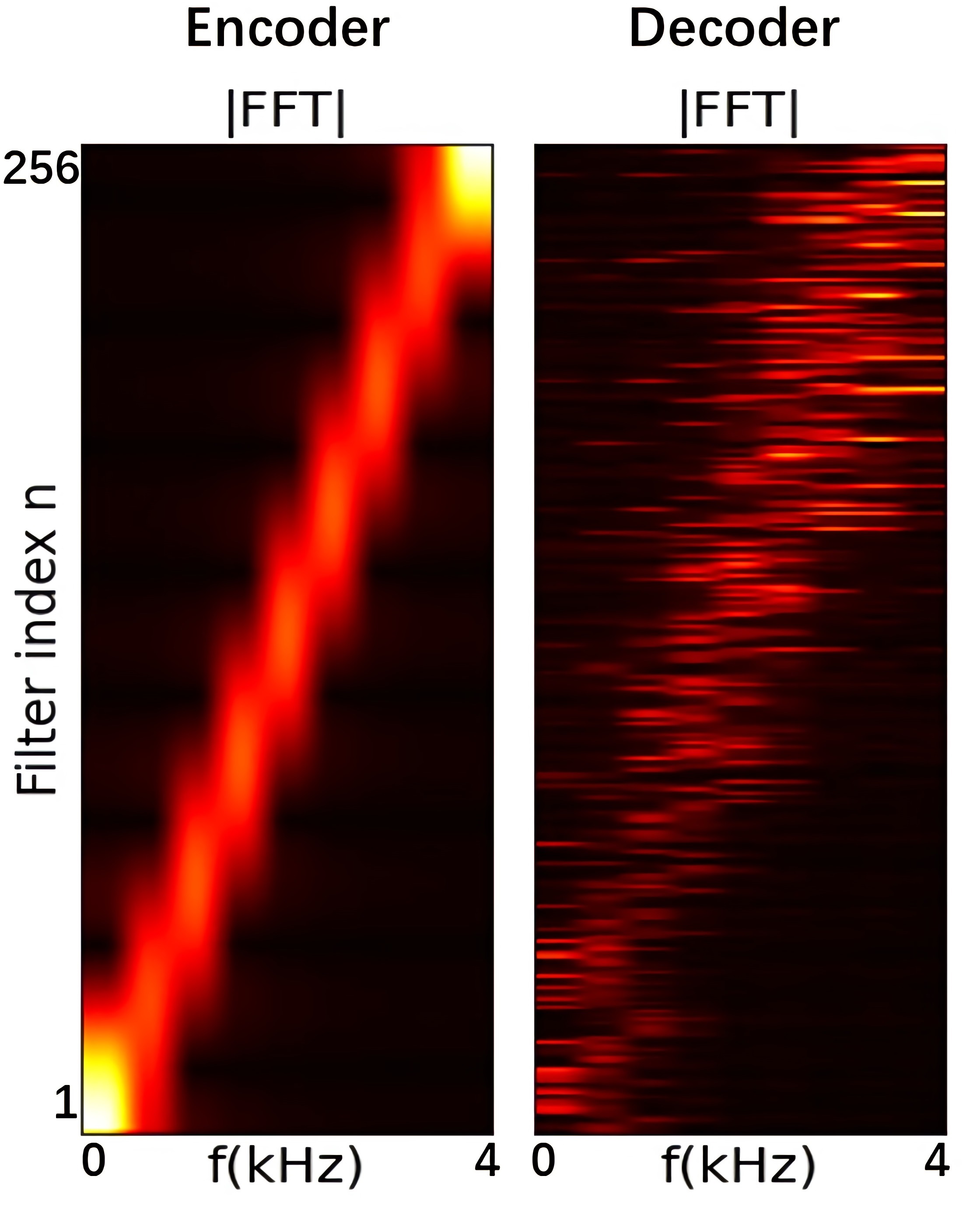}
\end{minipage}
}
\caption{Visualization of different configuration of encoder and learned decoder magnitudes of their FFTs. The left is a MPGTF-based encoder, the middle is a ParaMPGTF-based encoder and the right is a STFT-based encoder.}
\label{Weights_FFT_jpg}
\end{figure*}

\subsection{Results with learnable decoders}

We first conducted a comparison between STFT, MPGTF, ParaMPGTF, and learnable features when the decoders were set to the learnable features.
The comparison results are listed in Table \ref{table2}. From the table, we observe that the four features do not yield fundamentally different performance. If we look at the details, we find that STFT reaches the highest SI-SNR in both the development set and the test set. MPGTF and ParaMPGTF show competitive performance, where ParaMPGTF performs slightly better than MPGTF on the development set, and slightly worse than the latter on the test set.

\begin{table}[!t]
\centering
\caption{Comparison of different encoders when the decoders are set to learnable filters. }
\vspace{1mm}
\scalebox{0.95}{
\begin{tabular}{c|c|c|c|c}
\hline
\textbf{Encoder} & \textbf{Decoder}&\textbf{Mask activation} & \multicolumn{2}{c}\textbf{SI-SNR (dB)}\\
\hline & & & Dev&Test\\
\hline Learned & Learned & Sigmoid & 17.61&16.92\\
\hline Learned & Learned & ReLU &17.45&16.89\\
\hline MPGTF & Learned &ReLU &17.66&17.20\\
\hline ParaMPGTF & Learned & ReLU &17.71&17.06\\
\hline STFT & Learned & ReLU &\textbf{17.96}&\textbf{17.28}\\
\hline
\end{tabular}
}
\label{table2}
\end{table}

\begin{table}[!t]
\centering
\caption{Comparison of $c_{1}$ and $c_{2}$ between MPGTF and ParaMPGTF when the decoders are set to learnable features.}
\vspace{1mm}
\begin{tabular}{c|c|c}
\hline
 & \textbf{MPGTF}&\textbf{ParaMPGTF}\\
 \hline $c_{1}$ & 24.7 & 25.09\\
 \hline $c_{2}$ & 9.265 & 9.198\\
\hline
\end{tabular}
\label{table4}
\end{table}

Fig. \ref{Weights_FFT_jpg} shows the magnitude spectrograms of the MPGTF, ParaMPGTF, and STFT encoders with their corresponding learnable decoders, where we only plot the STFT bins with indices from $1$ to $256$ \cite{8701652,pandey2019new} since that the real and imaginary parts share similar patterns. The filters are uniformly distributed in the frequency range from $0$ Hz to $4000$ Hz.
 From the figure, we see that the magnitude spectrograms of ParaMPGTF and MPGTF are similar. This phenomenon not only accounts for their similar performance, but also demonstrates that the parameterized feature is able to be optimized successfully. As a byproduct, it shows that (i) MPGTF is a well-designed handcrafted feature; (ii) the learnable decoders are able to learn effective inverse transforms of their encoders.

 Table \ref{table4} lists the comparison between the handcrafted $c_1$ and $c_2$ in MPGTF and the optimized $c_1$ and $c_2$ in ParaMPGTF. From the table, we see that the two groups of the parameters are similar, which further accounts for the similar performance of MPGTF and ParaMPGTF.

\subsection{Results with (pseudo) inverse transform decoders}

In this experiment, we set the encoder to STFT, MPGTF, and ParaMPGTF respectively, and set the decoder to their inverse transforms accordingly.

Table \ref{table3} lists the performance of MPGTF, ParaMPGTF, and STFT with their (pseudo) inverse transforms. From the table, we see that the performance of the three comparison methods is similar in general. If we look into the details, we see that the proposed ParaMPGTF reaches the best performance among the comparison methods on both the development set and the test set, which demonstrates the potential of the parameterized training strategy in improving conventional handcrafted features.

\begin{table}[!t]
	\centering
	\caption{Comparison of encoders and decoders with different features. The mask activation funciton is ReLU.}
\vspace{1mm}
	\begin{tabular}{c|c|c|c}
		\hline\textbf{Encoder} & \textbf{Decoder} &\multicolumn{2}{c}\textbf{SI-SNR (dB)}\\
		\hline && Dev&Test\\
		\hline MPGTF & MPGTF Pseudo Inv.&16.32 & 15.73\\
		\hline ParaMPGTF & ParaMPGTF Pseudo Inv.&\textbf{16.64} & \textbf{16.04}\\
		\hline STFT & ISTFT &16.31&15.82\\
		\hline
	\end{tabular}
	\label{table3}
\end{table}

\begin{figure}
	\vspace{-6mm}
	\centering
	\includegraphics[width=0.46\textwidth]{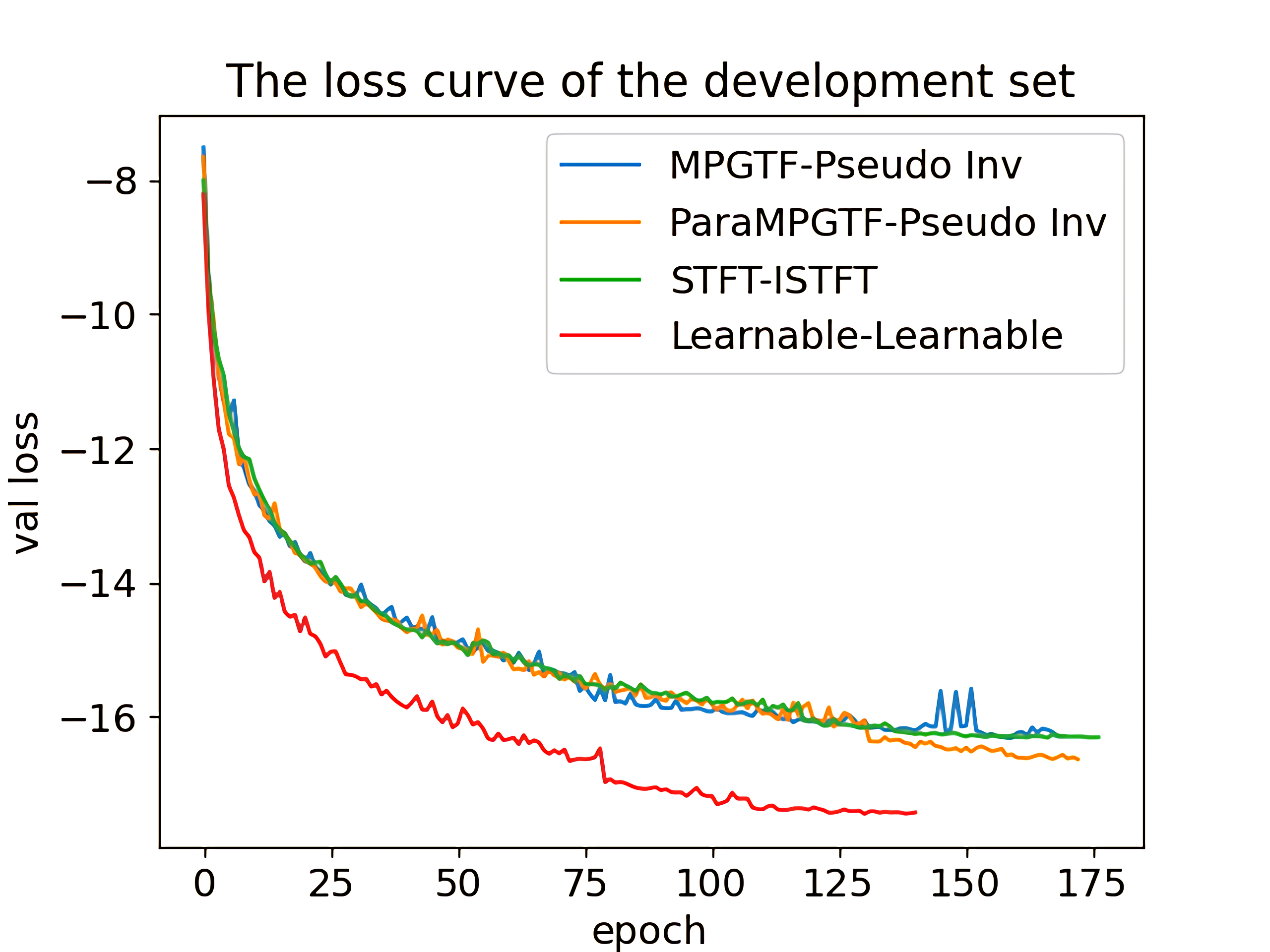}
	\caption{Convergence curves of different encoder-decoder pairs in the training process.}
	\label{val_loss_jpg}
\end{figure}

Fig. \ref{val_loss_jpg} shows the convergence curves of the deep models on the development set when the decoders are set to the (pseudo) inverse transforms of their encoders. From the figure, we find that the learnable feature converges faster than the handcrafted and parameterized features. Although
the handcrafted features and ParaMPGTF converge in a similar rate at the early training stage, ParaMPGTF converges faster at the late training stage.

\section{CONCLUSIONS}
\label{sec:conclusion}

In this paper, we have proposed a parameterized multi-phase gammatone filterbank. ParaMPGTF jointly learns the core parameters of MPGTF with the separation network. We have also compared handcrafted, parameterized, and learnable features in the same experimental framework, which is to our knowledge the first time that the three kinds of features are compared together, where the features in comparison are STFT, MPGTF, ParaMPGTF, and learnable features. Experiment results show that, when the decoders are set to learnable features, the four features behave similarly. STFT behave slightly better than the others. When the decoders are set to the (pseudo) inverse transforms of the encoders, ParaMPGTF performs better than the handcrafted features.

\bibliographystyle{IEEE}
\bibliography{myrefs}

\begin{thebibliography}{10}

\bibitem{8369155}
D.~{Wang} and J.~{Chen},
\newblock ``Supervised speech separation based on deep learning: An overview,''
\newblock {\em IEEE/ACM Transactions on Audio, Speech, and Language
  Processing}, vol. 26, no. 10, pp. 1702--1726, 2018.

\bibitem{7471631}
J.~R. {Hershey}, Z.~{Chen}, J.~{Le Roux}, and S.~{Watanabe},
\newblock ``Deep clustering: Discriminative embeddings for segmentation and
  separation,''
\newblock in {\em 2016 IEEE International Conference on Acoustics, Speech and
  Signal Processing (ICASSP)}, 2016, pp. 31--35.

\bibitem{7952154}
D.~{Yu}, M.~{Kolbæk}, Z.~{Tan}, and J.~{Jensen},
\newblock ``Permutation invariant training of deep models for
  speaker-independent multi-talker speech separation,''
\newblock in {\em 2017 IEEE International Conference on Acoustics, Speech and
  Signal Processing (ICASSP)}, 2017, pp. 241--245.

\bibitem{7979557}
M.~{Kolbæk}, D.~{Yu}, Z.~{Tan}, and J.~{Jensen},
\newblock ``Multitalker speech separation with utterance-level permutation
  invariant training of deep recurrent neural networks,''
\newblock {\em IEEE/ACM Transactions on Audio, Speech, and Language
  Processing}, vol. 25, no. 10, pp. 1901--1913, 2017.

\bibitem{7952155}
Z.~{Chen}, Y.~{Luo}, and N.~{Mesgarani},
\newblock ``Deep attractor network for single-microphone speaker separation,''
\newblock in {\em 2017 IEEE International Conference on Acoustics, Speech and
  Signal Processing (ICASSP)}, 2017, pp. 246--250.

\bibitem{1711.00541}
Yi~Luo and Nima Mesgarani,
\newblock ``Tasnet: time-domain audio separation network for real-time,
  single-channel speech separation,'' 2017.

\bibitem{1809.07454}
Yi~Luo and Nima Mesgarani,
\newblock ``Conv-tasnet: Surpassing ideal time-frequency magnitude masking for
  speech separation,''
\newblock 2018.

\bibitem{8683634}
A.~{Pandey} and D.~{Wang},
\newblock ``Tcnn: Temporal convolutional neural network for real-time speech
  enhancement in the time domain,''
\newblock in {\em ICASSP 2019 - 2019 IEEE International Conference on
  Acoustics, Speech and Signal Processing (ICASSP)}, 2019, pp. 6875--6879.

\bibitem{1902.04891}
Ziqiang Shi, Huibin Lin, Liu Liu, Rujie Liu, Jiqing Han, and Anyan Shi,
\newblock ``Furcanext: End-to-end monaural speech separation with dynamic gated
  dilated temporal convolutional networks,'' 2019.

\bibitem{9053602}
D.~{Ditter} and T.~{Gerkmann},
\newblock ``A multi-phase gammatone filterbank for speech separation via
  tasnet,''
\newblock in {\em ICASSP 2020 - 2020 IEEE International Conference on
  Acoustics, Speech and Signal Processing (ICASSP)}, 2020, pp. 36--40.

\bibitem{1910.10400}
Manuel Pariente, Samuele Cornell, Antoine Deleforge, and Emmanuel Vincent,
\newblock ``Filterbank design for end-to-end speech separation,'' 2019.

\bibitem{1808.00158}
Mirco Ravanelli and Yoshua Bengio,
\newblock ``Speaker recognition from raw waveform with sincnet,'' 2018.

\bibitem{10.1007/978-3-319-49409-8_7}
Colin Lea, Ren{\'e} Vidal, Austin Reiter, and Gregory~D. Hager,
\newblock ``Temporal convolutional networks: A unified approach to action
  segmentation,''
\newblock in {\em Computer Vision -- ECCV 2016 Workshops}, Gang Hua and
  Herv{\'e} J{\'e}gou, Eds., Cham, 2016, pp. 47--54, Springer International
  Publishing.

\bibitem{8099596}
C.~{Lea}, M.~D. {Flynn}, R.~{Vidal}, A.~{Reiter}, and G.~D. {Hager},
\newblock ``Temporal convolutional networks for action segmentation and
  detection,''
\newblock in {\em 2017 IEEE Conference on Computer Vision and Pattern
  Recognition (CVPR)}, 2017, pp. 1003--1012.

\bibitem{1992Complex}
R.~D. Patterson, K.~Robinson, J.~Holdsworth, D.~Mckeown, C.~Zhang, and
  M.~Allerhand,
\newblock ``Complex sounds and auditory images,''
\newblock {\em Auditory Physiology and Perception}, pp. 429--446, 1992.

\bibitem{2002Frequency}
V~Hohmann,
\newblock ``Frequency analysis and synthesis using a gammatone filterbank,''
\newblock {\em Acta Acustica United with Acustica}, vol. 88, no. 3, pp.
  433--442, 2002.

\bibitem{8701652}
A.~{Pandey} and D.~{Wang},
\newblock ``A new framework for cnn-based speech enhancement in the time
  domain,''
\newblock {\em IEEE/ACM Transactions on Audio, Speech, and Language
  Processing}, vol. 27, no. 7, pp. 1179--1188, 2019.

\bibitem{pandey2019new}
Ashutosh Pandey and DeLiang Wang,
\newblock ``A new framework for cnn-based speech enhancement in the time
  domain,''
\newblock {\em IEEE/ACM Transactions on Audio, Speech, and Language
  Processing}, vol. 27, no. 7, pp. 1179--1188, 2019.

\end{thebibliography}

\end{document}